# Coauthorship and Thematic Networks in AAEP Annual Meetings

Juan M.C. Larrosa

*jlarrosa@criba.edu.ar*

Departamento de Economía, Universidad Nacional del Sur (UNS)

Instituto de Investigaciones Económicas y Sociales del Sur (IIESS)

======== First Draft ===========

**Abstract**

We analyze the coauthorship production of the AAEP Annual Meeting since 1964. We use social network analysis for creating coauthorship networks and given that any paper must be tagged with two JEL codes, we use this information for also structuring a thematic network. Then we calculate network metrics and find main actors and clusters for coauthors and topics. We distinguish a gender gap in the sample. Thematic networks show a cluster of codes and the analysis of the cluster shows the preeminence of the tags related to trade, econometric, distribution/poverty and health and education topics.

## 1. Introduction

Networks of collaboration among researchers have been presented since early times of science. To co-author a paper means collaboration between at least two authors and is a way of learning, sharing knowledge, and labor division. We construct networks of scientists in which a link between two scientists is established by their coauthorship of one or more scientific papers. These networks are affiliation networks in which actors are linked by their common membership of groups consisting of authors of a paper. They are social networks with more interaction than many affiliation networks (networks where people are connected to an event, for instance); it is probably fair to say that most people who have written a paper together are genuinely acquainted with one another. We study many statistical properties of our networks, including numbers of papers written by coauthors, the timeline of papers presented, the topics that identify each contribution, the typical distance through the network from one scientist to another, and a variety of measures of connectedness within a network. Specifically, we estimate centralities such as closeness, eigenvector and betweennes.

These specific networks became popular when mathematicians calculate their own Erdös number by counting how many coauthors they share with any other coauthor of the prolific Paul Erdös. Those who had published a paper with him were given a Erdös number of 1, those who had published with one of those people but not with Erdös, a number of 2, and so on (Newmn). So far, studies of current literature about coauthorship networks mostly give emphasis to understand patterns of scientific collaborations, to capture collaborative statistics, and to propose valid and reliable measures for identifying prominent author(s).

Coauthorship network, as an important kind of social network, has been intensively studied (Newman (2001a, 2001b, 2003, 2005); Barabási et al. (2002); Nascimento et al. (2003); Kretschmer (2004); Liu et al. (2005); Yin et al. (2006); Vidgen et al. (2007); Rodriguez and Pepe (2008), Uddin et al. (2012), Rumsey (2006), Paredes (2011), Erfanmanesh et al. (2012), Kurosawa and Takama (2012), Brandao and Moro (2012), Murray et al. (2006), Jahn (2008), Maia et al. (2013), Murray et al. (2006), Day and Shih (2011)). A complex approach to diverse and enormous datasets of different branches of the science is presented in Cotta and Merelo (2005).

## 2. Coauthor networks

What is a co-authorship network? When two authors write a paper, they establish a link between them. As long as other authors collaborate in producing one paper more links are included relating each new co-author with all other previous authors. We study the way coauthorship between economists who present contributions in a periodical professional meeting in Argentina has been evolved in the past decades. The common place for exposing new contributions in the Economics profession in Argentina has been the traditional Annual Meeting of the Asociación Argentina de Economía Política (AAEP) where newly from recently graduated to old school researchers and professors converges yearly for presenting a highly diverse supply of contributions[1].

Figure 1 shows four examples or real coauthorship networks constructed with data of AAEP annual meetings. The central node of each (star-like) network is the named author. For instance, the coauthorship network of Leonardo Gasparini shows several lines connecting only two nodes (a paper of him and only one coauthor), triangles (a paper with two coauthors), and more complicated forms. In the case of Walter Sosa Escudero's coauthorship network shows also lineas, triangles, and even a heptagon. This is the case of the largest coauthored paper presented in the AAEP meetings with seven coauthors.

[1] Chinchilla-Rodríguez et al. (2012) study a more broad approach to Argentinian scientific authors but focusing on publications in indexed journals. Aguado-López et al. (2009) also find pattern of collaboration among coauthors in Argentina.

**Figure 1. Examples of AAEP coauthorship networks**

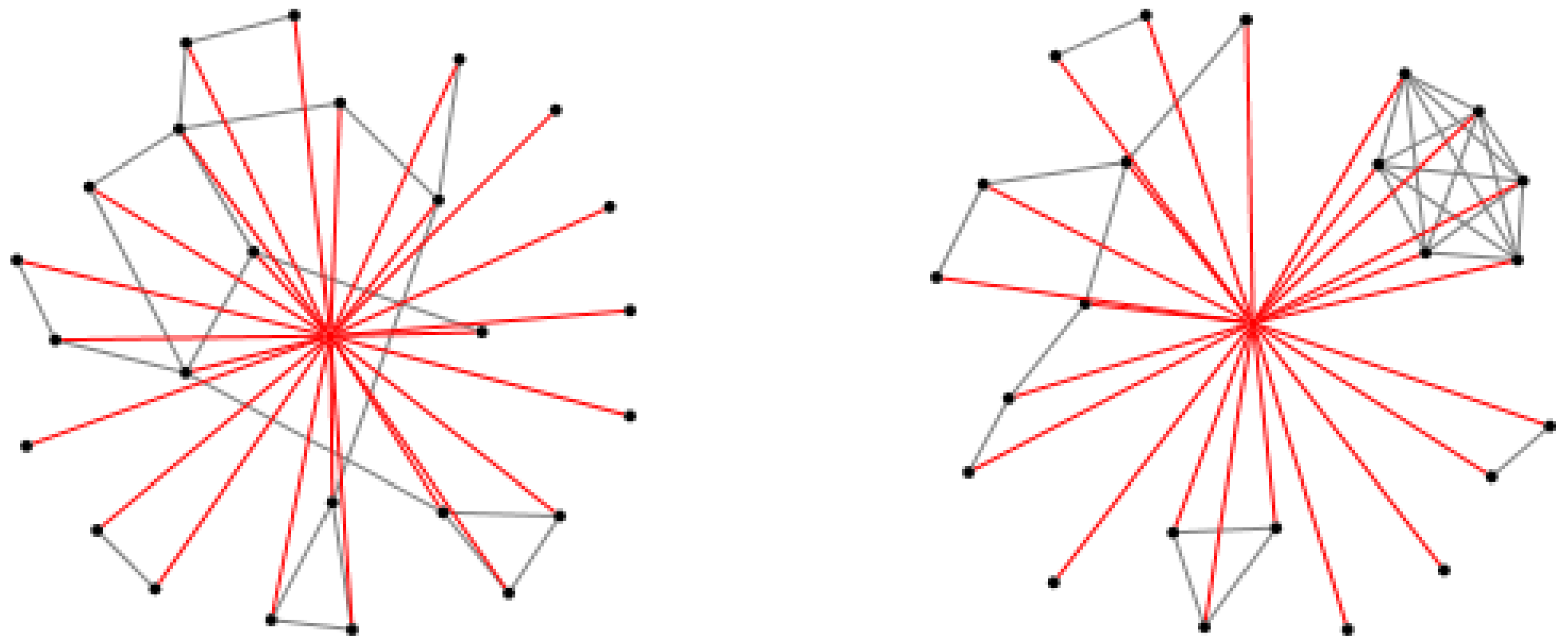

**Coauthorship network of Leonardo Gasparini** **Coauthorship network of Walter Sosa Escudero**

Note: Center of the star network is the named author. Source: The Author

Other contributions have added information on network of collaboration between economists. Fafchamps et al. (2006) and Goyal et al (2006) present evidence of small world properties in an enormous database of economist contributions. Bukowska et al. (2014) also present a specific study of coauthorship between Polish economists[2]. These three previous papers focus in structural properties of the network and focus on journal references. The first two conclude, as well as in our case, that the collaboration community of economists represents a small world framework, where most contributors belong to a huge connected component and they coexist with several isolated participants.

As shown in Figure 2, the quantity in coauthored contributions in the AAEP Annual Meeting has increased remarkably since the 1990s. Each of the networks depicted in the position upper to the bars represents a ten year window for coauthorship papers. The earliest (network above bars to the extreme left) is for the 60's, the second for the 70's, and so on. The last network represent the 5 year- period of 2010-2014. It is clear that networks become larger and more populated. The 90s marks a notorious increment and the last network also shows that the last five years presents a level of activity similar to previous decade in terms of coauthorship.

[2] Medical sciences have been prone to use this methodology. Valderrama-Zurián et al. (2007) uses coauthorship network analysis but for a Spanish cardiologist journal and Ramírez Ruiz (2009) for Spanish psychiatrist researchers. González-Alcaide and Navarro-Molina (2008) makes the same approach for reproductive biology's literature. Olmeda-Gómez et al. (2009) also study Spanish intra and interuniversity and communities coauthorship. Medicis Morel et al. (2009) uses coauthorship for planning purposes in disease prevention.

**Figure 2. Evolution in the number of coauthorship papers and network of coauthors**

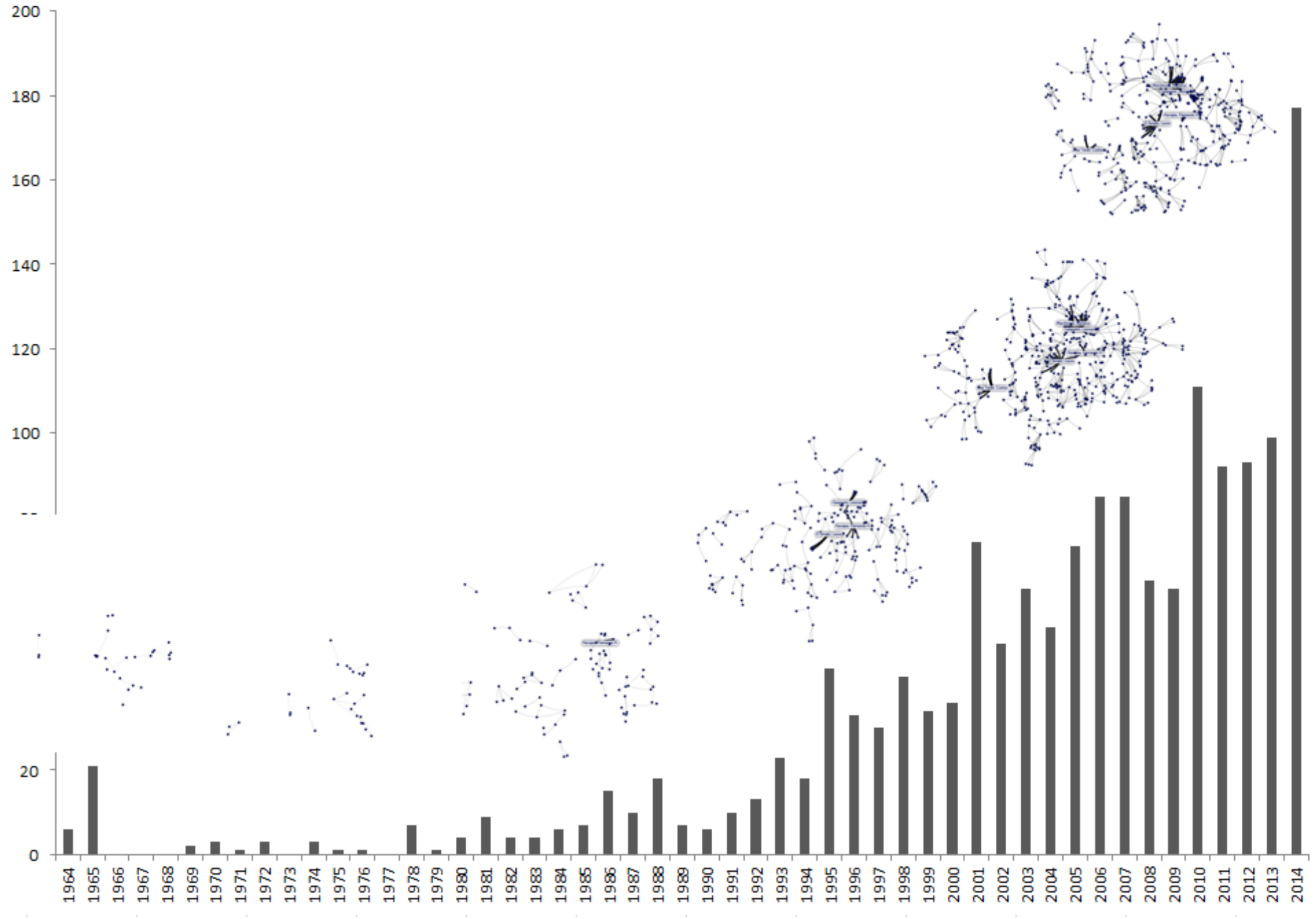

We want to explore the pattern of co-authorship networks and to describe the collaboration structure between authors, institutions and topics covered by the contributions. We will make use of metrics of social network analysis, from to a network level (structural) and to actor level. We use clustering (indirect approach), graph metrics (such as density, diameter, etc.), centrality indices; cohesion measures (component, cliques)[3]. A first glance at the data can be observed in Figure 3. The links of three big actors are enlarged in black.

[3] Choobdar et al. (2012) use a motif approach for detecting groups in coauthorship networks.

**Figure 3. Coauthorship network of AAEP Annual Meeting with its three main coauthors remarked (1964-2014)**

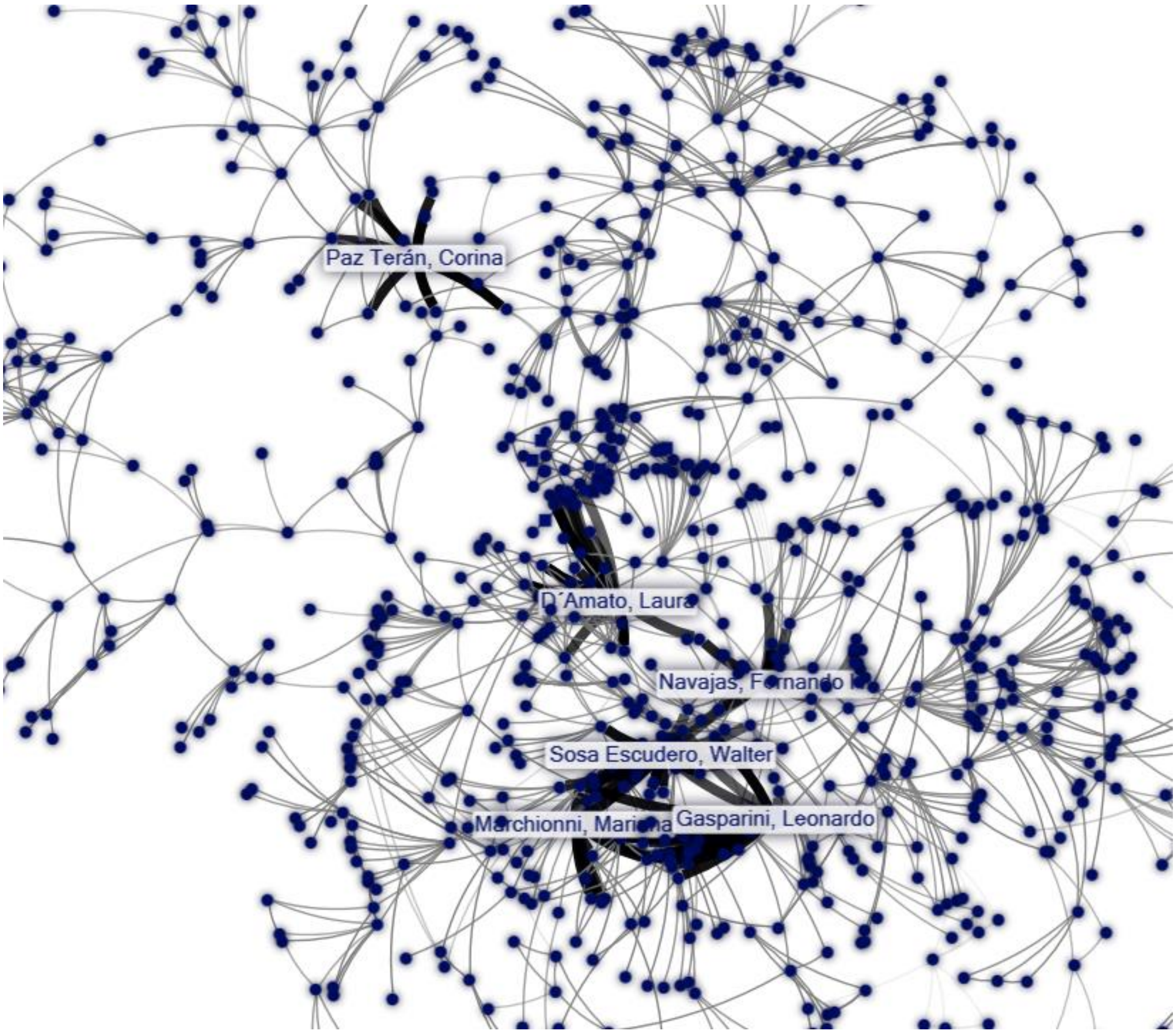

**Note: Six main actors labeled and their links are enlarged. Network layout is Harel-Koreln Fast Multiscale (HKFM)**

## 2.1 Methods

To undertake this study we identified conference papers published in the AAEP site during 1964-2014. For each paper selected we identified the name and surnames of the authors, as well as their institutional affiliation (institution) and one or two JEL codes of the paper. The network consists in 890 nodes and 1644 edges.

We obtained a series of measures to allow for the analysis of structure or social networks. If we take agents (authors or institutions) individually, we present 3 measures of centrality or cohesion that facilitate detailed analysis of the social network studied: degree, and indices of betweenness and closeness.

Several social network metrics are employed that enable to measure the characteristics of a network, including components analysis and centrality analysis. Explanations for component analysis and the metrics used are as follows.

### 2.1.1 Component Analysis

A component of a network is a substructure in which there is at least a path connecting a node and any other node. A network may include some components which are isolated from each other without any connections. The size of a component indicates the number of nodes it contains. Component analysis is used to study the network structure. The component analysis is employed to study academic circles of the AAEP proceedings in this paper.

### 2.1.2 Centrality Analysis

The earliest pursued category of methodology in the social network analysis is the centrality of individuals and organizations in their social networks. Different kinds of centralities including degree, betweenness and closeness, give rough indications of the social power from several different perspectives of a node based on how well they "connect" the network[4].

Degree centrality denotes the number of links connected to a node in the network. Important nodes usually have high degree. $C_D(n_i)$ is the degree of node $n_i$ and is calculated by the equation as follows.

$$C_D(n_i) = \sum_{i \neq j} X_{ij}$$

where $X_{ij}$ equals 0 or 1, 0 means actor *i* has no tie with actor *j*, while 1 means they do have a link.

The normalized value of degree centrality of actors in the network can be calculated by the formula as follows.

$$C'_D(n_i) = \sum_{i \neq j} \frac{X_{ij}}{g-1}$$

The *g* represents the total number of the actors in the network. Degree indicates in our case the number of different economists, a specific author or several other authors, that are directly connected with, and is obtained by identifying and subsequently quantifying relationships of coauthorship. It's a measure that reflects the greater or lesser extent of collaboration maintained by authors.

Betweenness indicates the extent to which a particular node lies between the various other nodes in the network. For instance, among 3 nodes, A, B and C, A is connected with C and B is connected with C, but no connection exists between A and B. Therefore, C is the key node between A and B when A and B want to connect to each other. The role of broker or gatekeeper is played by the actor with high betweenness who at same time is endowed a potential for control over others.

The betweenness centrality of a node in a network is calculated by the following formula.

---

[4] Abbasi et al. (2011) uses these metrics for correlating with coauthor personal academic performance.

$$C_B(n_i) = 2\sum_{j\prec k} g_{jk}(n_i)/g_{jk}$$

The normalized value of the betweenness centrality of a node in a network is calculated by the following formula.

$$C'_B(n_i) = 2\frac{\sum_{j\prec k} g_{jk}(n_i)/g_{jk}}{(g-1)(g-2)}$$

Betweenness determines the extent to which an agent is situated in the middle of or between other agents in the network, permitting us to make interconnections. Betweenness measures the prestige of authors and institutions and their capacity to access and control information flow. It is calculated as the sum of the shortest paths between the 2 agents that include between them the agent in question.

Eigenvector centrality is calculated by assessing how well connected an individual is to the parts of the network with the greatest connectivity. This information is derived by the eigenvalues of the eigenvector of the adjacency matrix of the network (matrix representation of the graph). Individuals with high eigenvector scores have many connections, and their connections, have many connections as well. The mathematical notion behind eigenvectors is that we can express the matrix of nodes' connections (adjacency) with a set of characteristic values that are assigned to each individual. Eigenvector centrality scores correspond to the score you get for individuals if you start by constructing the connections node-to-node between all individuals in a network, and then assign a single number to each individual while attempting to keep the distances between these new values equal to the observed distances. Of course this can't be done with a single numeric value per individual, but in fact you can always represent a set of social connections or distances by assigning as many vectors (strings of numbers for each individual) as there are individuals in the network.

High eigenvector centrality individuals are leaders of the network. They are often public figures with many connections to other high-profile individuals. Thus, they often play roles of key opinion leaders and shape public perception. However, these persons are not necessarily associated to another centrality measures (like betweenness and degree).

PageRank centrality outputs a probability distribution used to represent the likelihood that by randomly exploring on the links of a network one will arrive to any particular other author. PageRank works by counting the number and quality of links to an author (in our case) to determine a rough estimate of how important the author is. The underlying assumption is that more important authors are likely to receive more links from other coauthors.

**Figure 4. Total papers presented at the AAEP Annual Meeting (single and coauthored)**

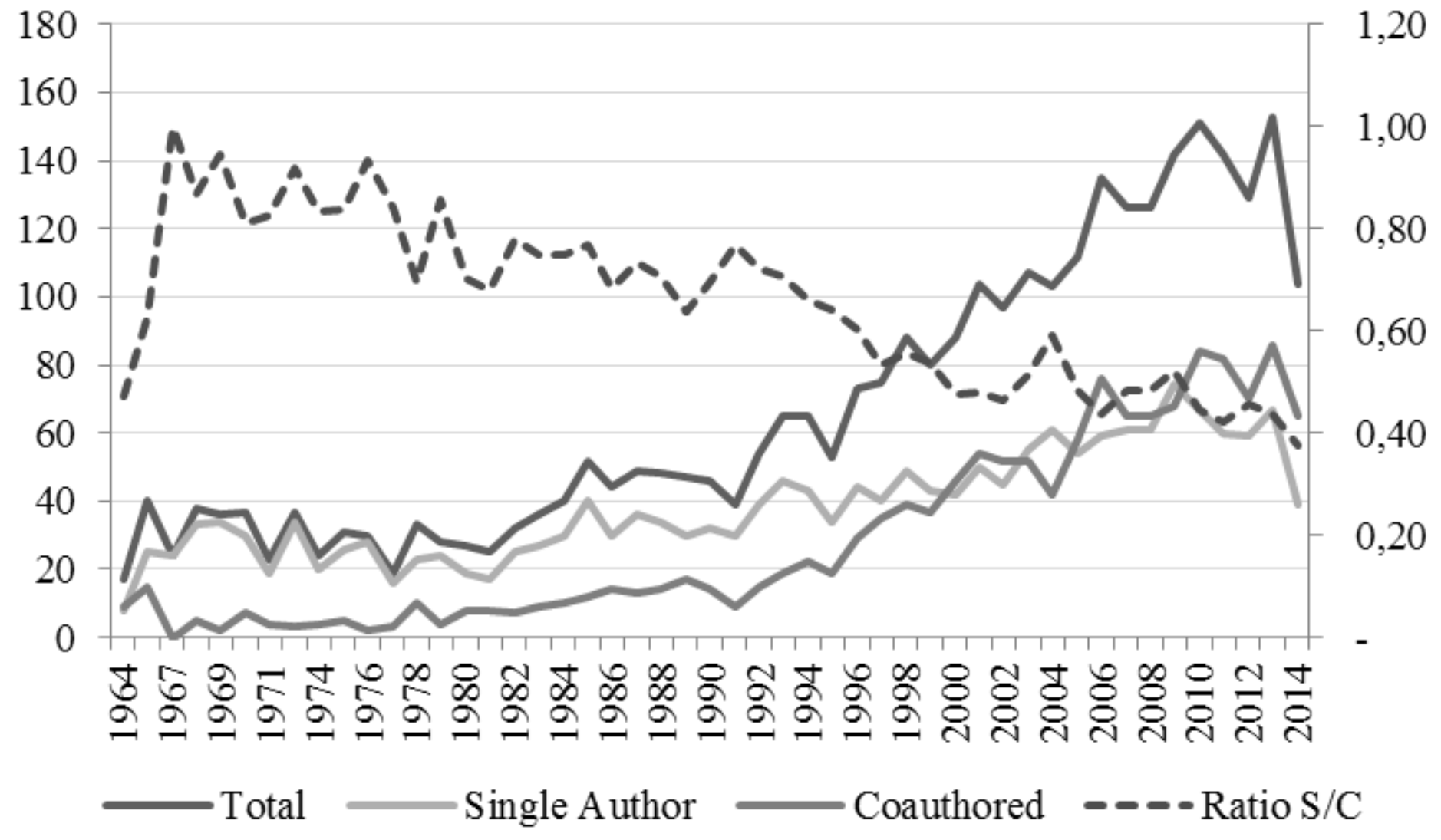

### 2.2 Descriptive Data of the Proceedings

As mentioned earlier, the AAEP is the Asociación Argentina de Economía Política (Argentinian Political Economics Association) and congregates most of the professional and academic economists in the country. It has been realized one annual meeting since 1964 in different universities and cities across the country. In 1966 and 1973 the meeting was suspended. As a byproduct of each meeting the association prints the proceedings of each meeting that are actually online.

Papers have been primarily single-authored with a decreasing trend. As Figure 4 depicts, most papers were single-authored initially up to mid-seventies but it with the beginning of the nineties where  Single authored papers have preeminent in the proceedings of the congress up to 1999 when coauthored contributions surpassed them in quantity.  As the ratio Single/Coauthored (S/C) shows in the dashed line coauthorship have been steadily increasing (S/C ratio decreasing) since the 1970s.

### 2.3 Small world

In graph theory, a connected component of an undirected graph (such as our case) is a subgraph (a network inside the network) in which any two vertices are connected to each other by paths (steps of links connecting nodes), and which is connected to no additional vertices in the supergraph (the rest of the network). The whole network has up to 17 connected components but the main component add up the 96% of nodes and 98% of links (see Table 1). This is a case of the small world phenomenon (SWP) with a big connected component that connects most of the nodes. A very simple measure of SWP is that the natural logarithm of the total of nodes (*ln*(890)=6.74) roughly approximates to the average distance among peers (8.45). The diameter of the larger connected component is high requiring as much as 20 steps for beginning with one node to reach its farthest node. The SWP is the observation that one can find a short chain of colleagues, often of no more than a handful of individuals, connecting almost any two authors on the network. It is equivalent to the statement that most pairs of individuals are connected by a short path through the coauthorship network.

**Table 1. Graph Metrics of Connected Components**

| Group | N | UE | EwD | TE | MGD | AGD | D |
|---|---|---|---|---|---|---|---|
| G1 | 850 | 1442 | 167 | 1609 | 20 | 8,45 | 0,004 |
| G2 | 5 | 10 | 0 | 10 | 1 | 0,8 | 1 |
| G3 | 3 | 3 | 0 | 3 | 1 | 0,667 | 1 |
| G4 | 3 | 3 | 0 | 3 | 1 | 0,667 | 1 |
| G5 | 3 | 3 | 0 | 3 | 1 | 0,667 | 1 |
| G6 | 3 | 3 | 0 | 3 | 1 | 0,667 | 1 |
| G7 | 3 | 3 | 0 | 3 | 1 | 0,667 | 1 |
| G8 | 2 | 1 | 0 | 1 | 1 | 0,5 | 1 |
| G9 | 2 | 1 | 0 | 1 | 1 | 0,5 | 1 |
| G10 | 2 | 1 | 0 | 1 | 1 | 0,5 | 1 |
| G11 | 2 | 1 | 0 | 1 | 1 | 0,5 | 1 |
| G12 | 2 | 1 | 0 | 1 | 1 | 0,5 | 1 |
| G13 | 2 | 1 | 0 | 1 | 1 | 0,5 | 1 |
| G14 | 2 | 1 | 0 | 1 | 1 | 0,5 | 1 |
| G15 | 2 | 1 | 0 | 1 | 1 | 0,5 | 1 |
| G16 | 2 | 1 | 0 | 1 | 1 | 0,5 | 1 |
| G17 | 2 | 1 | 0 | 1 | 1 | 0,5 | 1 |

**Codification: N = Nodes; UE = Unique edges; EwD = Edges with duplicates; TE = Total edges; MGD = Maximum geodesic distance (diameter); AGD = Average geodesic distance; D = Graph density** (Source: The Author)

Another descriptive metric that is interesting to remark is presented in the composite Figure 5. The main figure shows the distribution frequency of the degree among coauthors and the inserted figure represents the same information but in log-log scale. As it’s observed, it is a long tail distribution of degrees (fat-tail distribution). This is an indication for the presence of a structure of diffusion of information congregated in hubs, or persons with many connections that reach farther nodes. A network with fat-tail distribution of degree of nodes has good properties for dissemination of information.

**Figure 5. Frequency of Degrees and Log-log Representation**

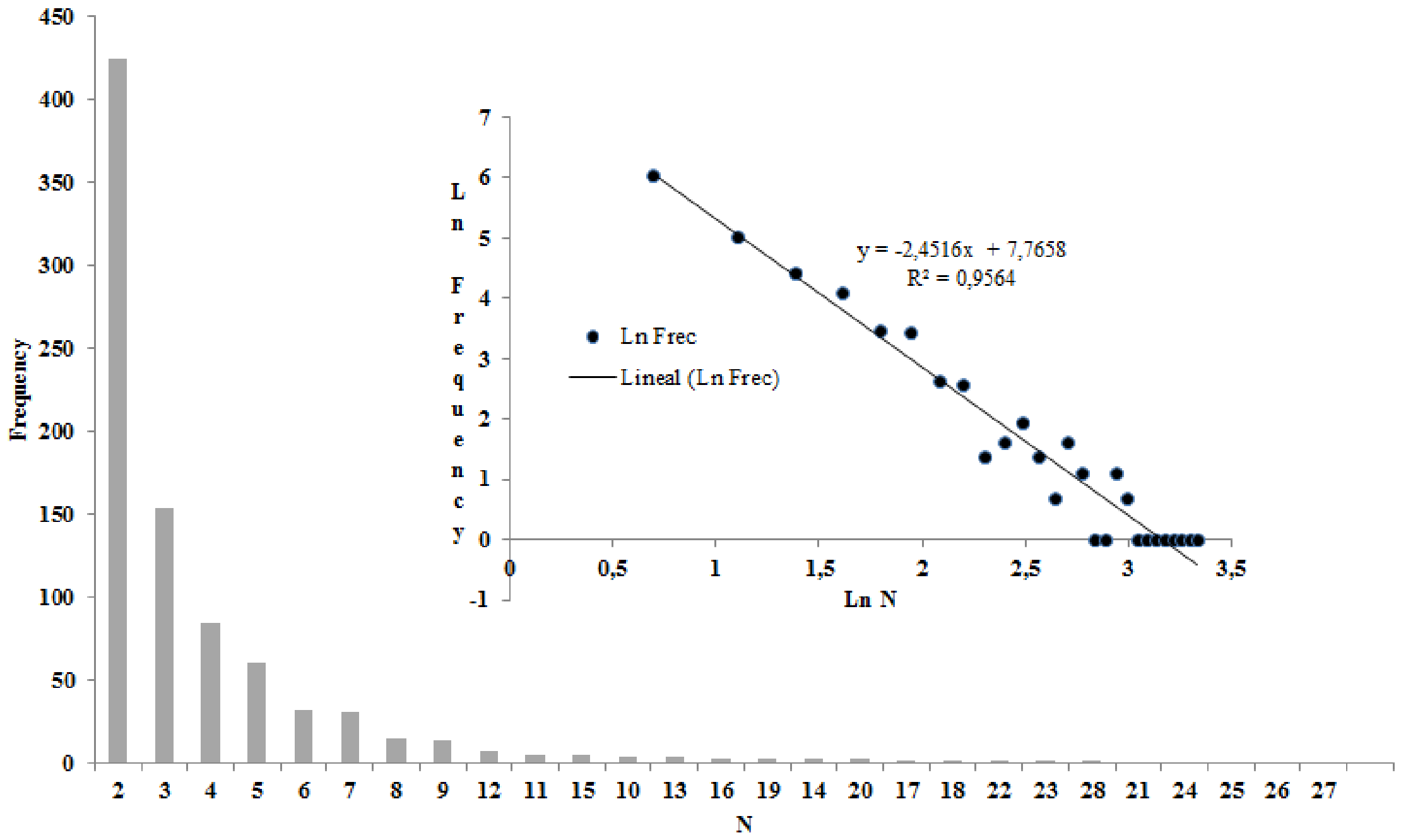

## 2.4 Affiliations

Members of the economic community in Argentina work in diverse public and private institutions. These were classified into 19 national universities (Universidad Nacional de Córdoba-UNC-, Universidad Nacional de La Plata –UNLP-, Universidad Nacional del Sur –UNS-, Universidad de Buenos Aires –UBA-, Universidad Nacional de Cuyo –UNCu-, Universidad Nacional de Salta –UNSa-, Universidad Nacional de Rosario –UNR-, Universidad Nacional de Mar del Plata –UNMdP-, Universidad Nacional General Sarmiento –UNGS-, Universidad Nacional del Litoral –UNL-, Universidad Nacional de Río Cuarto –UNRC-, Universidad Nacional de Chilecito -UN Chilecito-, Universidad Nacional de la Patagonia San Juan Bosco –UNPSJB-, Universidad Nacional de San Luis –UNSL-, Universidad nacional de Morón –UNaM-, Universidad Nacional del Centro de la Provincia de Buenos Aires –UNCPBA-, Universidad Nacional de Quilmes –UNQ-, Universidad Nacional de Tucumán –UNSTA-, Universidad Nacional de San Martín –UNSAM-), 9 private universities (Universidad del CEMA –UCEMA-, Universidad de San Andrés –UDESA-, Universidad Torcuato Di Tella –UTDT-, Universidad Argentina de la Empresa –UADE-, Universidad Católica Argentina –UCA-, Universidad Austral –Uaustral-, Universidad de Belgrano –UB-, Universidad Empresarial Siglo 21 -UES21-, Universidad Católica de Santa Fé –UCSF-), several Private/Public Research Center, Foreign Universities, State Entities (ministries and the like), Private/Public Bank, International Entities (UN, CEPAL, among others). Figure 6 shows up the clusters detected by affiliation. Clusters are presented in circular network layout and the thinner links represent intra-group coauthorship and broader links represent extra-group coauthorship. The size of each circle network makes it easy to grasp what affiliations are more prominent than other.

**Figure 6. Cluster by affiliation**

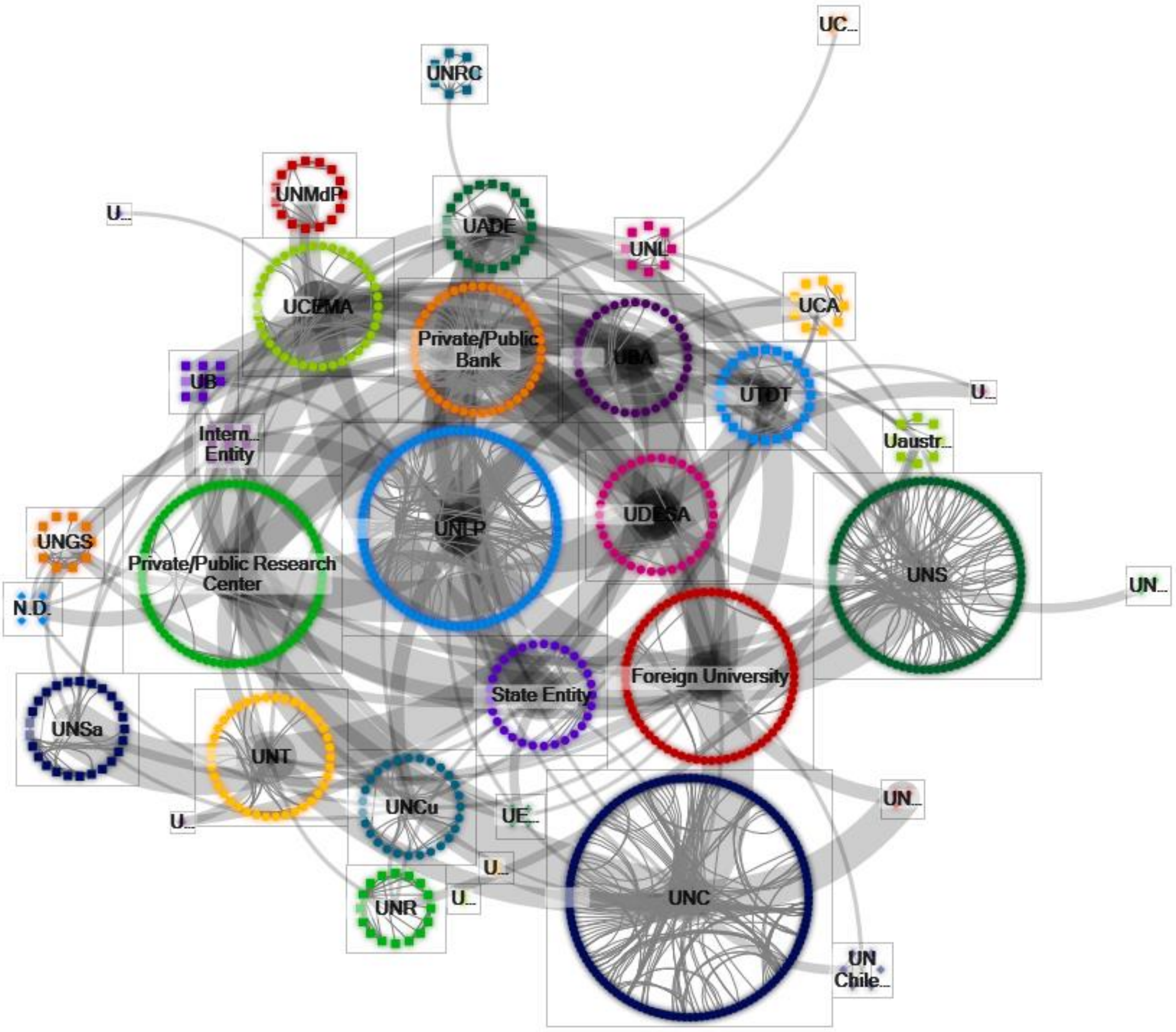

### 2.4.1 Structural metrics

As Table 2 shows up the larger affiliation that engenders coauthorship to the meetings is UNC with 130 coauthors followed in sharp difference by the UNL and UNS, 42 and 44 coauthors below respectively.

**Table 2. Cluster metrics of groups with more than 20 nodes**

| Label | N | UE | EwD | TE | CC | SVCC | MVCC | MECC | MGD | AGD | D |
|---|---|---|---|---|---|---|---|---|---|---|---|
| **UNC** | 130 | 226 | 46 | 272 | 4 | 2 | 125 | 269 | 10 | 4,23 | 0,030 |
| **UNLP** | 88 | 116 | 17 | 133 | 12 | 9 | 73 | 126 | 8 | 3,54 | 0,032 |
| **UNS** | 86 | 155 | 6 | 161 | 3 | 2 | 84 | 161 | 8 | 3,85 | 0,043 |
| **Foreign University** | 63 | 33 | 0 | 33 | 39 | 28 | 7 | 9 | 3 | 1,08 | 0,017 |
| **Public/Private Research** | 57 | 31 | 0 | 31 | 31 | 20 | 7 | 7 | 4 | 1,22 | 0,019 |
| **Central/Private Bank** | 41 | 50 | 18 | 68 | 11 | 7 | 27 | 64 | 5 | 2,23 | 0,071 |
| **UCEMA** | 38 | 38 | 2 | 40 | 12 | 9 | 21 | 31 | 7 | 2,88 | 0,055 |
| **UDESA** | 34 | 17 | 0 | 17 | 20 | 14 | 8 | 10 | 4 | 1,60 | 0,030 |
| **UNT** | 34 | 40 | 2 | 42 | 8 | 5 | 25 | 40 | 8 | 3,37 | 0,073 |
| **State Entity** | 29 | 20 | 2 | 22 | 16 | 10 | 5 | 7 | 2 | 0,81 | 0,052 |
| **UBA** | 29 | 17 | 0 | 17 | 15 | 10 | 7 | 9 | 2 | 1,21 | 0,042 |
| **UNCu** | 27 | 30 | 0 | 30 | 6 | 4 | 21 | 29 | 8 | 3,29 | 0,085 |
| **UTDT** | 24 | 13 | 4 | 17 | 12 | 9 | 11 | 15 | 5 | 2,14 | 0,054 |
| **UNSa** | 22 | 28 | 4 | 32 | 2 | 1 | 21 | 32 | 5 | 2,57 | 0,130 |

**Codification: N = Nodes; UE = Unique edges; EwD = Edges with duplicates; TE = Total edges; CC = Connected components; SVCC = Single-Vertex connected components; MVCC = Maximum vertices in a connected component; MECC = Maximum edges in a connected component; MGD = Maximum geodesic distance (diameter); AGD = Average geodesic distance; D = Graph density (Source: The Author)**

After metrics have been estimated we obtain the correlations between them. Table 3 shows the correlation matrix. The size of the cluster (N) is highly correlated with the presence of unique edges (UE) and total edges (TE) for one side, and maximum vertices (MVCC) and edges (MECC) in a connected component, for the other side. The five metrics are logically correlated with the size of the cluster given that larger number of coauthors requires more links and vertices. Density (D) is slightly negative correlated with almost all metrics except with those related to diameter (MGD, AGD). Edges with duplicates (EwD), that represent repeated coauthorships, are also related with the size (N, TE, UE, MVCC, MECC) of the cluster,

**Table 3. Correlations among cluster metrics**

| | N | UE | EwD | TE | CC | SVCC | MVCC | MECC | MGD | AGD | D |
|---|---|---|---|---|---|---|---|---|---|---|---|
| **N** | 1 | | | | | | | | | | |
| **UE** | 0,935 | 1 | | | | | | | | | |
| **EwD** | 0,780 | 0,838 | 1 | | | | | | | | |
| **TE** | 0,930 | 0,996 | 0,885 | 1 | | | | | | | |
| **CC** | -0,062 | -0,402 | -0,373 | -0,406 | 1 | | | | | | |
| **SVCC** | -0,062 | -0,402 | -0,383 | -0,408 | 0,995 | 1 | | | | | |
| **MVCC** | 0,890 | 0,988 | 0,842 | 0,986 | -0,508 | -0,502 | 1 | | | | |
| **MECC** | 0,893 | 0,990 | 0,883 | 0,995 | -0,490 | -0,489 | 0,993 | 1 | | | |
| **MGD** | 0,567 | 0,714 | 0,572 | 0,707 | -0,618 | -0,597 | 0,777 | 0,742 | 1 | | |
| **AGD** | 0,538 | 0,732 | 0,569 | 0,721 | -0,736 | -0,710 | 0,808 | 0,769 | 0,974 | 1 | |
| **D** | -0,525 | -0,267 | -0,147 | -0,252 | -0,632 | -0,633 | -0,168 | -0,179 | 0,124 | 0,224 | 1 |

Source: The Author

# 3. Network of Coauthors

When analyzing important actors in the coauthorship sample we estimated actor's centralities. Gender differences emerge. Figure 7 shows the difference in size in male vs. female coauthorship network.

**Figure 7. Male and Female Coauthorship Sub-Networks**

**a) Male coauthorship network** **b) Female coauthorship network**

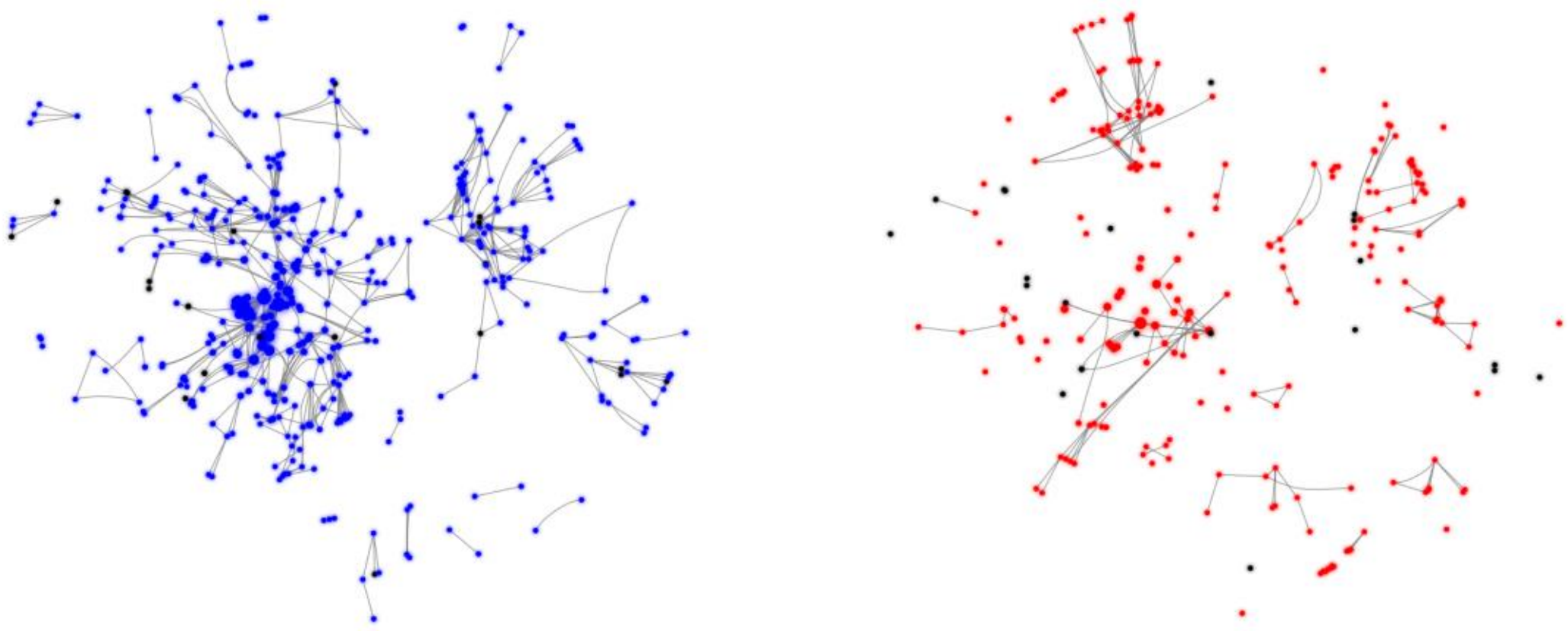

Source: The Author

A way to notice another gender differences we can anticipate that most centrality calculations sensitively differ when considering male or female coauthorship subnetwork[5]. For instance, Table 4 shows the deviations from the mean in centralities between genders. Male coauthors surpass female colleagues in all centrality measures. Female, on the other hand, shows a slightly upper estimation for clustering coefficient. That would mean women tend to join papers with more coauthors.

[5] Yan et al. (2009) use centrality measures for estimating the impact of their publication.

**Table 4. Mean Deviations of each Metric by Gender**

| | Degree | Betweenness Centrality | Closeness Centrality | Eigenvector Centrality | PageRank | Clustering Coefficient |
|---|---|---|---|---|---|---|
| Male | 3,1% | 18,2% | 5,0% | 18,9% | 4,4% | -6,0% |
| Female | -6,4% | -37,0% | -10,9% | -38,6% | -8,9% | 12,3% |

Sample: Female: 279 individuals, Male: 566 individuals. Source: The Author

Before getting into individual actor analysis, we can observe in Figure 8 a matrix of scatterplots between estimated full sample centralities. At first glance, PageRank and degree show a remarkably correlation with a less clear image in the case of betweenness and PageRank and degree. That is to say, actors with high degree tend to have high PageRank and in some cases high betweenness.

**Figure 8. Matrix of Scatterplots of Coauthor Centralities**

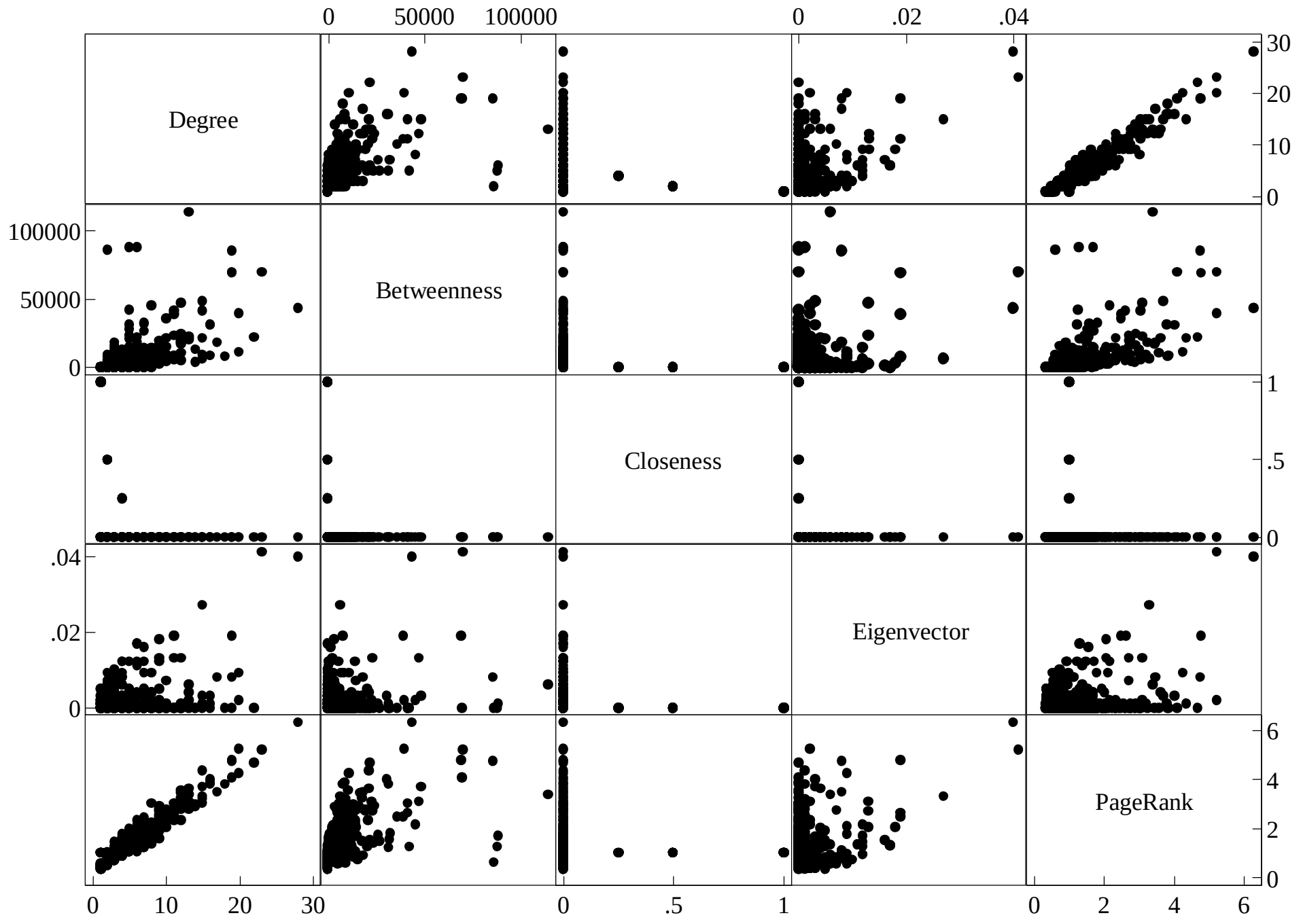

Having shown this difference among economists by gender, it is plausible to see more male economists in the top of centrality measures. The following tables will remark this fact. In the top ten considering the entire sample, given the gender gap presented in Table 4, we will observe mostly male economist. So Table 6 to Table 8 show maxima from sample and for male economists at the same time (with some exceptions visible in the table). Table 5 shows up actors with the larger betweenness centrality. Fernando Navajas and Santiago Urbiztondo (both from UCEMA). This centrality remarks economists that tend to connect groups otherwise disconnected.

**Table 5. Top Ten Betweenness Centrality Authors**

| Author | Degree | Betweenness | Affiliation |
|---|---|---|---|
| **Navajas, Fernando H.** | 13 | 114033,23 | FIEL |
| Urbiztondo, Santiago | 6 | 88394,89 | UCEMA |
| Abdala, Manuel Angel | 5 | 87869,3 | UNC |
| Spiller, Pablo T. | 2 | 86203,3 | UDESA |
| Heymann, Daniel | 19 | 85440,2 | UBA |
| **Sosa Escudero, Walter** | 23 | 69798,03 | UDESA |
| Arrufat, José Luis | 19 | 69434,26 | UNC |
| Porto, Alberto | 19 | 68620,18 | UNLP |
| Tohmé, Fernando | 15 | 48178,99 | UNS |
| Canavese, Alfredo Juan | 12 | 46950,28 | UTDT |

In the cases of degree and eigenvector centralities, Table 6 shows that Leonardo Gasparini (UNLP) has the maximum degree and has the second maximum Walter Sosa Escudero (UNLP and UDESA). This order is reversed when considering eigenvector centrality, where Sosa Escudero has the maximum followed by Gasparini. Higher degree points to coauthor having worked with more other coauthors and eigenvector points out to coauthor having worked with more important (in terms of degree) other coauthors.

**Table 6. Top Ten Eigenvector Centrality Authors**

| Author | Degree | Eigenvector | Affiliation |
|---|---|---|---|
| **Sosa Escudero, Walter** | 23 | 0,041 | UDESA |
| **Gasparini, Leonardo** | 28 | 0,04 | UNLP |
| Marchionni, Mariana | 15 | 0,027 | UNLP |
| Porto, Alberto | 19 | 0,019 | UNLP |
| Streb, Jorge | 11 | 0,019 | UCEMA |
| Druck, Pablo | 11 | 0,019 | Fundación del Tucumán |
| Rodriguez, Carlos A. | 9 | 0,018 | UCEMA |
| Bolzico, Javier | 6 | 0,017 | UCEMA |
| Henke, Alejandro | 6 | 0,017 | UCEMA |
| Rutman, José | 6 | 0,017 | BCRA |

PageRank centrality metric highlights authors that are more solicited by other coauthors for producing contributions. Popular authors emerge clearly with this metric, and again, Gasparini (UNLP) and Sosa Escudero (UDESA) jointly with Carrera (UNLP-BCRA) seem to be the most popular coauthors in the event.

**Table 7. Top Ten PageRank Centrality Authors**

| Author | Degree | PageRank | Affiliation |
|---|---|---|---|
| **Gasparini, Leonardo** | 28 | 6,289 | UNLP |
| Carrera, Jorge | 20 | 5,227 | UNLP |
| **Sosa Escudero, Walter** | 23 | 5,221 | UDESA |
| Porto, Alberto | 19 | 4,783 | UNLP |
| Heymann, Daniel | 19 | 4,766 | UBA |
| Gertel, Héctor | 22 | 4,689 | UNC |
| Galperín, Carlos | 15 | 4,372 | UBA |
| D´Amato, Laura | 20 | 4,269 | BCRA |
| Arrufat, José Luis | 19 | 4,089 | UNC |
| Dabús, Carlos | 16 | 4,004 | UNS |

It is fair to remark that Sosa Escudero is the coauthor that emerges as the only economist in the Top 10 of all centralities as observed in all the aforementioned tables.

Now let's focus on the gender gap. Tables 9 to 12 show the centrality metrics for female coauthor subnetwork. Specifically, Table 8 shows the top ten female economists sorted by degree. Heading the list are two BCRA economists: Laura D'Amato and Tamara Burdisso.

**Table 8. Top Ten Degree Centrality Female Coauthors**

| Author | Degree | Affiliation |
|---|---|---|
| **D´Amato, Laura** | 20 | BCRA |
| Burdisso, Tamara | 17 | BCRA |
| Marchionni, Mariana | 15 | UNLP |
| Moscoso, Nebel | 15 | UNS |
| London, Silvia | 12 | UNS |
| Picardi de Sastre, Marta Susana | 12 | UNS |
| Cerro, Ana María | 12 | UNT |
| Paz Terán, Corina | 11 | UNT |
| Conte Grand, Mariana | 9 | UCEMA |
| Recalde, Maria Luisa | 9 | UNC |

Table 9 shows a different picture in the top ten of the women economist having the highest betweenness centrality metrics. Corina Paz Terán from UNT and Iris Perlbach de Maradona (UNCu) are first and second, respectively. It is interesting to note that more than half of the main female brokers in the network are from universities of the interior of the country. That would mean that women tend to connect groups geographically disconnected.

**Table 9. Top Ten Betweenness Centrality Female Coauthors**

| Vertex | Degree | Betweenness Centrality | Affiliation |
|---|---|---|---|
| **Paz Terán, Corina** | 11 | 41.468 | UNT |
| Perlbach de Maradona, Iris | 7 | 32.301 | UNCu |
| Sarquis, Liliana | 7 | 31.609 | UNC |
| Juárez de Perona, Hada | 5 | 31.321 | UNC |
| London, Silvia | 12 | 23.854 | UNS |
| Navarro, Ana Inés | 5 | 23.124 | UNR |
| D' Elia, Vanesa | 6 | 21.328 | ANSES |
| Conte Grand, Mariana | 9 | 19.200 | UCEMA |
| Burdisso, Tamara | 17 | 17.797 | BCRA |
| Cerro, Ana María | 12 | 16.327 | UNT |

In Table 10 we present results from eigenvector centrality estimation. This time, emerge Mariana Marchionni (UNLP) and Hildegart Ahumada (UTDT) as the two more important nodes. It is worthy to note that Ahumada has less than half that Marchionni's eigenvector centrality and, as noted earlier, Marchionni has only the 65% of the same metric compared to maximum male economist. All of the female economists in the top 10 are affiliated to institution from the Buenos Aires Province.

**Table 10. Top Ten Eigenvector Centrality Female Coauthors**

| Author | Degree | Eigenvector Centrality | Affiliation |
|---|---|---|---|
| **Marchionni, Mariana** | 15 | 0,027 | UNLP |
| Ahumada, Hildegart | 9 | 0,012 | UTDT |
| D´Amato, Laura | 20 | 0,009 | BCRA |
| Alzúa, María Laura | 7 | 0,009 | UNLP |
| Gabrielli, María Florencia | 8 | 0,009 | BCRA |
| Burdisso, Tamara | 17 | 0,008 | BCRA |
| Serio, Monserrat | 4 | 0,008 | UNLP |
| Edo, María | 3 | 0,007 | UCEMA |
| Viollaz, Mariana | 4 | 0,006 | UNLP |
| Conconi, Adriana | 3 | 0,006 | CEDLAS |

Finally, Table 11 displays the top ten female economists according to PageRank centrality. This metric is also highly correlated with degree in the female subnetwork, so it is unsurprisingly similar to Table 8. D'Amato and Burdisso are the most central economist given by this metric but it is interesting to note that more than half of the top ten are economist from the interior of the country.

**Table 11. Top Ten PageRank Centrality Female Coauthors**

| Author | Degree | PageRank | Affiliation |
|---|---|---|---|
| **D´Amato, Laura** | 20 | 4,269 | BCRA |
| Burdisso, Tamara | 17 | 3,46 | BCRA |
| Cerro, Ana María | 12 | 3,441 | UNT |
| Marchionni, Mariana | 15 | 3,293 | UNLP |
| Moscoso, Nebel | 15 | 3,19 | UNS |
| London, Silvia | 12 | 2,921 | UNS |
| Conte Grand, Mariana | 9 | 2,707 | UCEMA |
| Paz Terán, Corina | 11 | 2,626 | UNT |
| Bergés, Miriam | 9 | 2,45 | UNMdP |
| Picardi de Sastre, Marta Susana | 12 | 2,352 | UNS |

## 4. Network of JEL categories

Each co-authored contribution is classified according to, at least, one or two JEL codes. This facilitates analyzing what topics is co-present in the economic community that co-authored papers in the AAEP event. We link each paper by the coincidence of its two JEL codes this way providing a picture of the topics that are shared by the contributions.

The thematic network has 109 nodes and 451 unique links and 34 self-loop links. The density of the graph is 0.071 and exhibit one connected component. The diameter of the network is 6 with an average geodesic distance of 2.6.

We calculate the random walk closeness and betweenness centralities because of the presence of self-loop in the data (Newman (2005)). Actually both centralities are highly correlated (correlation of .98). Figure 9 show the JEL code network. Clusters were made by grouping contributions within each general category (only first letter). Clusters are displayed in circular network. Slimmer links represent intra-cluster links and broader links represent extra-cluster links.

**Figure 9. Clustering by JEL category**

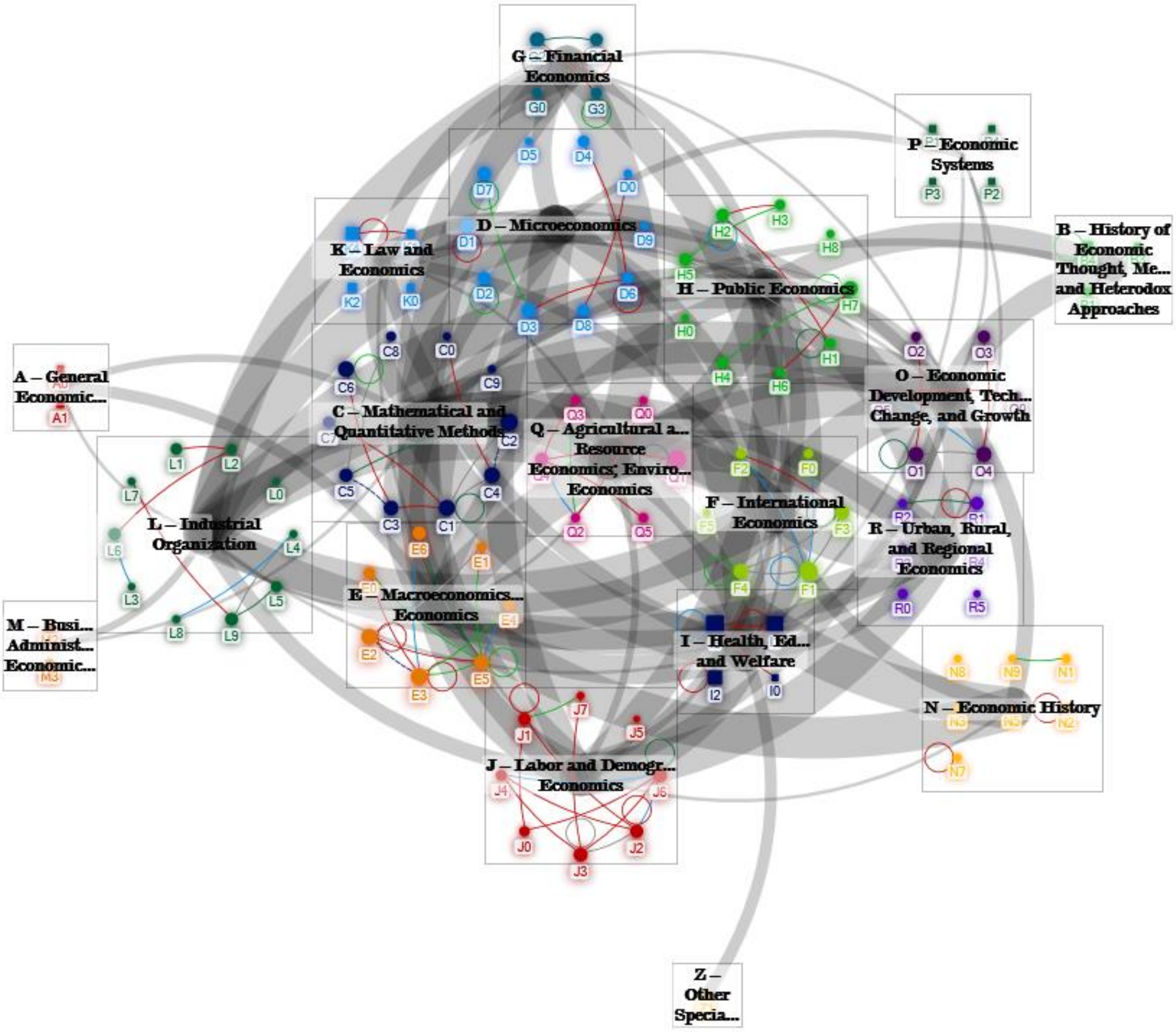

Again as in the previous section a descriptive metric about the diffusion property of the network is presented in the composite Figure 10. The main figure shows the distribution frequency of the degree among coauthors and the inserted figure represents the same information but in log-log scale. As observed, it appears to have a long tailed distribution of degrees as in Figure 5. This is too an indication for the presence of a structure of diffusion of information congregated in hubs, or persons with many connections that reach farther nodes. A network with fat-tail distribution of degree of nodes has good properties for dissemination of information.

**Figure 10. Frequency of Degrees and Log-log Representation**

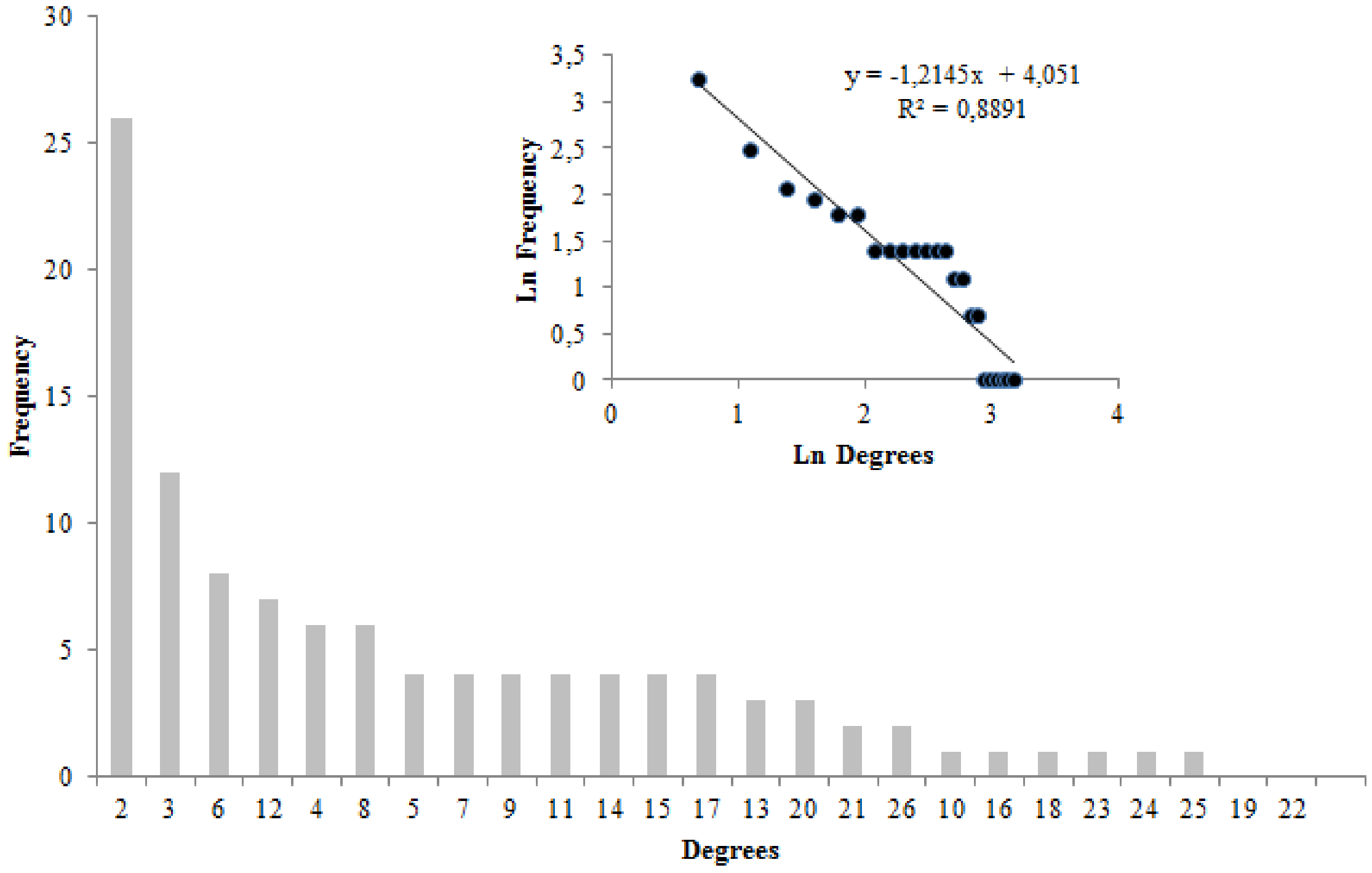

Network metrics of the cluster by one-letter JEL code network are presented in Table 12. The three larger clusters (higher N) correspond to C (Mathematical and Quantitative Methods), D (Microeconomics) and L (Industrial Organization). The higher density (D) in E (Macroeconomics and Monetary Economics) implies that macroeconomists tend to repeat the code. The highest diameter (MGD) in C (Mathematical and Quantitative Methods) and O (Economic Development, Technological Change, and Growth) implies that both JEL codes are related to other far diverse JEL codes across contributions. A (General Economics, Handbooks and Teaching), M (Business Administration and Business Economics; Marketing; Accounting), P (Economic Systems), and Z (Other Special Topics) are isolated codes barely used by authors.

**Table 12. Network metrics for JEL codes network**

| JEL Code | N | TE | SL | CC | SVCC | MVCC | MECC | MGD | AGD | D |
|---|---|---|---|---|---|---|---|---|---|---|
| **A** | 2 | 0 | 0 | 2 | 2 | 1 | 0 | 0 | - | - |
| **B** | 3 | 2 | 1 | 2 | 1 | 2 | 1 | 1 | 0,40 | 0,33 |
| **C** | **10** | 13 | 3 | 2 | 1 | **9** | 13 | **4** | **1,93** | 0,22 |
| **D** | **10** | 8 | **4** | **6** | 4 | 4 | 5 | 3 | 1,00 | 0,09 |
| **E** | 7 | 14 | 3 | 1 | 0 | 7 | 14 | 2 | 1,27 | **0,52** |
| **F** | 6 | 10 | **4** | 2 | 1 | 5 | 10 | 2 | 1,12 | 0,40 |
| **G** | 4 | 5 | 3 | 2 | 1 | 3 | 5 | 2 | 0,89 | 0,33 |
| **H** | 9 | 11 | 3 | 3 | 2 | 7 | 11 | 3 | 1,63 | 0,22 |
| **I** | 4 | 6 | 3 | 1 | 0 | 4 | 6 | 2 | 1,13 | 0,50 |
| **J** | 8 | **15** | **4** | 2 | 1 | 7 | **15** | 2 | 1,27 | 0,39 |
| **K** | 4 | 2 | 1 | 3 | 2 | 2 | 2 | 1 | 0,50 | 0,17 |
| **L** | **10** | 7 | 0 | 3 | 1 | 5 | 4 | 3 | 1,37 | 0,16 |
| **M** | 2 | 0 | 0 | 2 | 2 | 1 | 0 | 0 | - | - |
| **N** | 7 | 3 | 2 | **6** | **5** | 2 | 1 | 1 | 0,33 | 0,05 |
| **O** | 6 | 6 | 1 | 1 | 0 | 6 | 6 | **4** | 1,72 | 0,33 |
| **P** | 4 | 0 | 0 | 4 | 4 | 1 | 0 | 0 | - | - |
| **Q** | 6 | 7 | 1 | 1 | 0 | 6 | 7 | 3 | 1,44 | 0,40 |
| **R** | 6 | 2 | 1 | 5 | 4 | 2 | 2 | 1 | 0,50 | 0,07 |
| **Z** | 1 | 0 | 0 | 1 | 1 | 1 | 0 | 0 | - | - |

**Labels: A – General Economics, Handbooks and Teaching; B – History of Economic Thought, Methodology, and Heterodox Approaches; C – Mathematical and Quantitative Methods; D – Microeconomics; E – Macroeconomics and Monetary Economics; F – International Economics; G – Financial Economics; H – Public Economics; I – Health, Education, and Welfare; J – Labor and Demographic Economics; K – Law and Economics; L – Industrial Organization; M – Business Administration and Business Economics; Marketing; Accounting; N – Economic History; O – Economic Development, Technological Change, and Growth; P – Economic Systems; Q – Agricultural and Natural Resource Economics; Environmental and Ecological Economics; R – Urban, Rural, and Regional Economics; Z – Other Special Topics.**

**Codification: N = Nodes; SL = Self-Loops; TE = Total edges; CC = Connected components; SVCC = Single-Vertex connected components; MVCC = Maximum vertices in a connected component; MECC = Maximum edges in a connected component; MGD = Maximum geodesic distance (diameter); AGD = Average geodesic distance; D = Graph density (Source: The Author)**

The correlation table of these variables is showed in Table 13. It shows that the size of the cluster (N) is not only highly correlated with the number of edges (TE), maximum number of vertices (MVCC) and diameter (MGD). This means that clusters with high number of JEL code links end up connecting more other JEL code contributions. The high correlation between self-loops (SL) and total edges (TE) take into account the repetitiveness of collaborations between same authors. Density (D), on the other hand, is negatively correlated with single vertex components (SVCC) and positively correlated with maximum edges in component, in both cases in an obvious manner.

**Table 13. Correlation table for JEL codes network metrics**

| | *N* | *TE* | *SL* | *CC* | *SVCC* | *MVCC* | *MECC* | *MGD* | *AGD* | *D* |
|---|---|---|---|---|---|---|---|---|---|---|
| **N** | 1,00 | | | | | | | | | |
| **TE** | 0,75 | 1,00 | | | | | | | | |
| **SL** | 0,55 | 0,78 | 1,00 | | | | | | | |
| **CC** | 0,37 | -0,19 | 0,08 | 1,00 | | | | | | |
| **SVCC** | 0,10 | -0,40 | -0,10 | 0,94 | 1,00 | | | | | |
| **MVCC** | 0,75 | 0,93 | 0,61 | -0,29 | -0,49 | 1,00 | | | | |
| **MECC** | 0,64 | 0,98 | 0,76 | -0,31 | -0,47 | 0,93 | 1,00 | | | |
| **MGD** | 0,78 | 0,73 | 0,51 | -0,12 | -0,39 | 0,87 | 0,67 | 1,00 | | |
| **AGD** | 0,75 | 0,84 | 0,57 | -0,26 | -0,51 | 0,95 | 0,82 | 0,96 | 1,00 | |
| **D** | 0,22 | 0,67 | 0,58 | -0,56 | -0,73 | 0,64 | 0,70 | 0,53 | 0,66 | 1,00 |

**Codification: N = Nodes; SL = Self-Loops; TE = Total edges; CC = Connected components; SVCC = Single-Vertex connected components; MVCC = Maximum vertices in a connected component; MECC = Maximum edges in a connected component; MGD = Maximum geodesic distance (diameter); AGD = Average geodesic distance; D = Graph density (Source: The Author)**

Trade (F1) and Welfare and Poverty (I3) are jointly the most chosen JEL codes (higher degree). However, Economic Development (O1) is the pivotal topic of the contributions having both maximum centrality in random walk closeness and betweenness followed by Trade. The code associated to most referenced codes is Econometric Methods (C2) with the higher value in eigenvector centrality.

**Table 14. Centralities the JEL Codes of Coauthored Papers (sorted by degree)**

| **Code-Area** | **Degree** | **RW-Betweenness Centrality** | **RW-Closeness Centrality** | **Eigenvector Centrality** | **PageRank** |
|---|---|---|---|---|---|
| F1 – Trade | **26** | 24,701 | 4,01 | 3,0 | **2,895** |
| I3 – Welfare and Poverty | **26** | 9,814 | 1,3 | 3,0 | 2,889 |
| C2 – Econometric methods: Single equation models; Single variables | 25 | 8,87 | 1,27 | **3,4** | 2,608 |
| Q1 – Agriculture | 24 | 7,362 | 0,97 | 2,4 | 2,749 |
| E3 – Prices, Business Fluctuations, and Cycles | 23 | 12,068 | 1,87 | 3,1 | 2,37 |
| C1 – Econometric and Statistical Methods: General | 21 | 14,708 | 2,31 | 2,5 | 2,226 |
| I1 – Health | 21 | 9,999 | 1,63 | 2,0 | 2,421 |
| C4 – Econometric and Statistical Methods: Special Topics | 20 | 15,084 | 2,21 | 2,3 | 2,23 |
| C6 – Mathematical Methods; Programming Models; Mathematical and Simulation Modeling | 20 | 12,829 | 2,01 | 2,2 | 2,164 |
| E2 – Macroeconomics: Consumption, Saving, Production, Employment, and Investment | 20 | 12,072 | 1,83 | 2,6 | 2,013 |
| O4 – Economic Growth and Aggregate Productivity | 18 | 8,114 | 1,3 | 2,1 | 2,113 |
| C3 – Econometric Methods: Multiple or Simultaneous Equation Models | 17 | 23,75 | 2,84 | 2,3 | 1,81 |
| G2 – Financial institutions and Services | 17 | 6,797 | 1,12 | 2,2 | 1,747 |
| H7 – State and Local Government; Intergovernmental Relations | 17 | 0,956 | 0,16 | 1,6 | 2,019 |
| O1 – Economic development | 17 | **26,196** | **2,87** | 2,0 | 1,782 |
| F4 – Macroeconomic Aspects of International Trade and Finance | 16 | 3,778 | 0,58 | 2,2 | 1,634 |
| E5 – Monetary Policy, Central Banking, and the Supply of Money and Credit | 15 | 0,391 | 0,07 | 2,1 | 1,485 |
| I2 – Education | 15 | 14,515 | 2,05 | 1,8 | 1,563 |

**Bold figures represent maximum values in each variable. RW-Betweenness Centrality means random walk betweenness and RW-Closeness means random walk closeness. Eigenvector centrality is the original metric multiplied by 100.**

The analysis of the interaction between the centralities published in Table 14 was made by observing scatter plots *viz a viz* each pair of centralities. A general positive relation among all centralities is observed, in some cases clearer than in other, as seen in Figure 11.

**Figure 11. Matrix of Scatter Plots of JEL Codes Centralities**

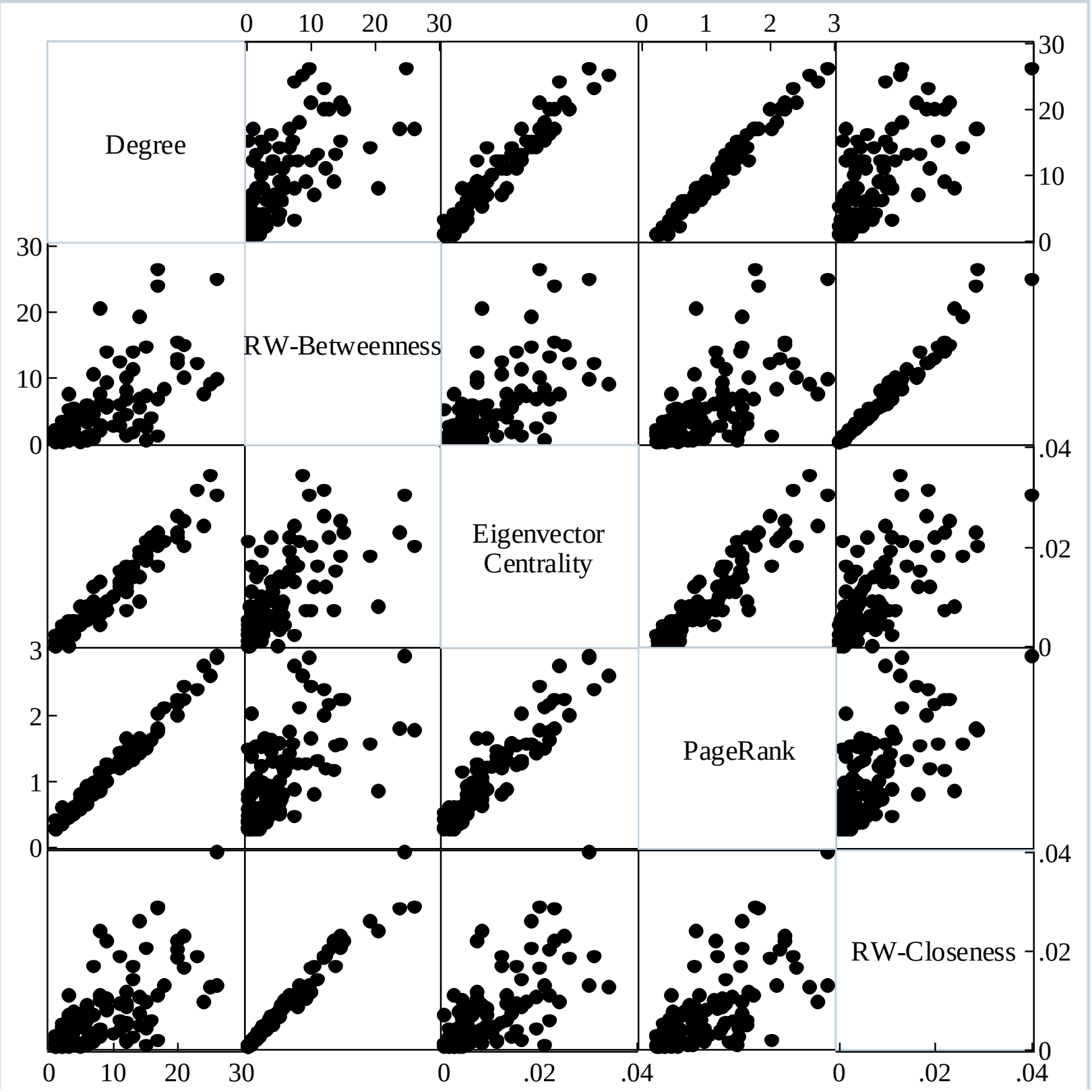

All this information is summarizes in the correlation matrix of Table 15. Random walk betweenness and closeness show high correlation meaning that codes more regularly used are close one to the other and they are also the codes that unifies otherwise separate clusters of topics.  On the other hand, degree and PageRank and eigenvector also show high correlation, meaning that highly used topics are also more popular directly and indirectly.

**Table 15. Correlation matrix of centralities**

| | *Degree* | *RW-Betweenness* | *Eigenvector* | *PageRank* | *RW-closeness* |
|---|---|---|---|---|---|
| Degree | 1 | | | | |
| RW-Betweenness | 0,655 | 1 | | | |
| Eigenvector | 0,966 | 0,657 | 1 | | |
| PageRank | 0,991 | 0,641 | 0,938 | 1 | |
| RW-closeness | 0,683 | 0,982 | 0,685 | 0,671 | 1 |

## 4.1 K-cores

A *k*-core in an undirected network is a connected maximal induced subgraph which has minimum degree greater than or equal to *k*. Moreover in our case, a *k*-core is a maximal group of JEL codes, all of whom are connected to some number (*k*) of other codes of the group. The coreness score is the maximum value of *k* for which it is in a *k*-core. In our case, the 7-core is the maximum value for *k*. It means that it takes to sever 7 links for any of the members of the core for disconnecting it off the group. Figure 12 shows the subgraph representing the 7-core. The red links are for the four JEL codes with the highest eigenvector centrality: C2 (Econometric methods: Single equation models; Single variables), E3 (Prices, Business Fluctuations, and Cycles), I3 (Welfare and Poverty), and F1 (Trade).

**Figure 12. 7-core of the network of JEL Codes**

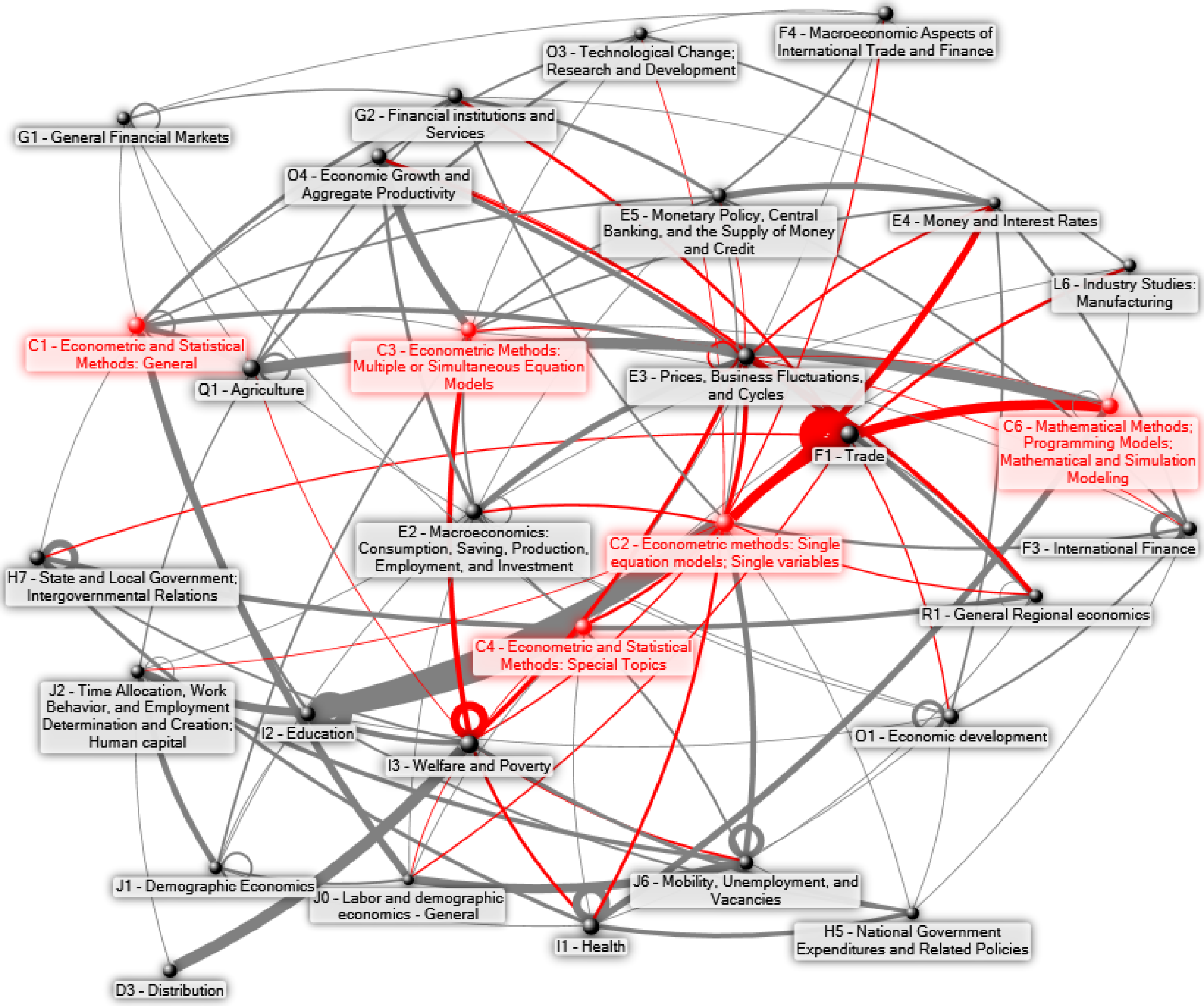

In Table 16 we publish the description of the codes for a better intuition about the relatedness of the connections. If we focus on Figure 12 we defined that broader links represent repeated interactions. In this case, at a glance, it seems that C2 and C4 interacts repeatedly with I2. So studies on Education have been supported by empirical evidence analyzed by econometric methods. It also can be noticed that Q1 and C6 are tagged jointly several times, grouping Agriculture studies with mathematical methods. C3 and O4 (with I3) are also jointly marked, given the idea that econometric methods also support evidence on growth poverty studies. A more powerful relation is show between more obviously I3 and D3 where poverty and distributional studies come together.  F1 are often related to C2, C6 and Q1, meaning that trade studies have been supported by econometrics and many times focused on agricultural markets. Summarizing, this 7-core indicates a group of highly interconnected topics that are frequently and jointly tagged among the contributions presented in the meeting.

**Table 16. 7-core of JEL codes description**

| Code |
|---|
| C1 – Econometric and Statistical Methods: General |
| C2 – Econometric methods: Single equation models; Single variables |
| C3 – Econometric Methods: Multiple or Simultaneous Equation Models |
| C4 – Econometric and Statistical Methods: Special Topics |
| C6 – Mathematical Methods; Programming Models; Mathematical and Simulation Modeling |
| D3 – Distribution |
| E2 – Macroeconomics: Consumption, Saving, Production, Employment, and Investment |
| E3 – Prices, Business Fluctuations, and Cycles |
| E4 – Money and Interest Rates |
| E5 – Monetary Policy, Central Banking, and the Supply of Money and Credit |
| F1 – Trade |
| F3 - International Finance |
| F4 – Macroeconomic Aspects of International Trade and Finance |
| G1 – General Financial Markets |
| G2 – Financial institutions and Services |
| H5 – National Government Expenditures and Related Policies |
| H7 – State and Local Government; Intergovernmental Relations |
| I1 – Health |
| I2 – Education |
| I3 – Welfare and Poverty |
| J0 – Labor and demographic economics - General |
| J1 – Demographic Economics |
| J2 – Time Allocation, Work Behavior, and Employment Determination and Creation; Human capital |
| J6 – Mobility, Unemployment, and Vacancies |
| L6 – Industry Studies: Manufacturing |
| O1 – Economic development |
| O3 – Technological Change; Research and Development |
| O4 – Economic Growth and Aggregate Productivity |
| Q1 – Agriculture |
| R1 – General Regional economics |

Source: The Author

## 5. Conclusions

We have presented data and analysis of coauthorship related to the state of the Economics profession in Argentina. We focused our effort in an annual congress where the majority of the community of economists in the country sends its production regularly. In this paper, we pave the way to study the impact of certain measures, such as centralities, densities, degree distributions, gender, location and affiliation in the production of economic contributions. The general features of the networks suggest promotes diffusion and several actors emerge as hubs of prestige. A thematic network is constructed by using the JEL codes (at most two for paper). We find a cluster of repeatedly tagged codes of 30 JEL Codes up to the first digit. It shows the importance of econometrics and mathematic models a supportive discipline of other more theoretical areas such agricultural economics, distributive and educational studies, trade and economic growth and development.

This is a first approach to the data and we think it opens a gate for study how the discipline has been evolved in Argentina at the main national event.